\documentstyle[amsfonts,aps]{revtex}

\begin{document}
\title{Internal Time Superoperator for Quantum Systems with Diagonal Singularity}
\author{Roberto Laura}
\address{Departamento de F\'{\i}sica\\
F.C.E.I.A.- Universidad Nacional de Rosario\\
Av. Pellegrini 250, 2000 Rosario, Argentina.\\
e-mail: laura@ifir.ifir.edu.ar}
\author{Adolfo R. Ord\'{o}\~{n}ez}
\address{Departamento de Matem\'{a}tica\\
F.C.E.I.A.- Universidad Nacional de Rosario\\
Av. Pellegrini 250, 2000 Rosario, Argentina.\\
e-mail: ordoniez@unrctu.edu.ar}
\date{September 10th.,1996.}
\maketitle

\begin{abstract}
We generalize the concepts of Internal Time Superoperator, its associated
non unitary similarity transformations and Liapounov variables, to quantum
systems with diagonal singularity, and we give a constructive proof of the
existence of these superoperators for systems with purely diagonal
Hamiltonian having uniform absolutely continuous spectrum on $[0,\infty ).$
\end{abstract}

\section{Introduction}

Recently, I. Antoniou et al.\cite{B}\cite{C}\cite{prep} have shown that
there is a natural formalism to deal with quantum mechanical systems with
continuous spectrum, as it is the case for decaying processes and non
equilibrium statistical mechanics. This formalism is in the line of an old
paper of Segal \cite{A} who assumed that a state $\rho $ is a positive
normal functional over the space of observables. The mean value of an
observable $O$ in the state $\rho $ is given by $\langle O\rangle _\rho
=\rho (O)$. This assumption implies an extension of the set of possible
states. This set is ''too small'' in the usual model, formulated in a
Hilbert space of density operators, to include the ''final'' state in the
case of {\em continuous spectrum} \cite{B}\cite{C}\cite{prep}. The new
formalism introduces the concept of {\em ''diagonal''} and {\em ''off
diagonal''} states and observables, with its corresponding projectors. For
this reason, the natural generalization consists of considering a *-algebra
of observables splited from the very beginning in two separated direct
summands. This, in turn, generates a corresponding spliting of the dual
space of states, as well as two projectors on the direct summands. What kind
of *-algebra of observables shall we take is not clear ''a priori''. It
could be a von Neumann or other C*-algebra, or a Banach algebra with
involution, or some other structure \cite{B}\cite{C}\cite{prep}.

The aim of this work is twofold. First we are going to give a natural
generalization of the concept of Internal Time Superoperator which measures
the {\em ''dynamical age''} or {\em ''degree of evolution''} of the system 
\cite{D}\cite{E}\cite{F}, its associated non unitary similarity
transformations and Liapunov Variables. Then we shall prove the existence of
these generalized superoperators for any quantum systems with a diagonal
Hamiltonian having uniform absolutely continuous spectrum on $[0,\infty )$.

In section 2, we begin with a simple introduction to the formalism of
quantum systems with diagonal singularity and introduce the necessary
notation.

The generalization of the time superoperator is given in section 3, together
with its spectral decomposition.

In section 4 we obtain a set of Liapounov variables.

In section 5 we show that our results prove the quantum analog of a well
known result: the existence of an Internal Time for classical K-fluxes \cite
{D}\cite{E}\cite{F}\cite{G}

\section{States and observables with diagonal singularity.}

Let us consider a quantum system such that, in the ordinary formalism, has a
Hamiltonian with a uniform absolutely continuous spectrum \cite{H} on $%
[0,\infty )$. For simplicity , we only consider the non degenerate case.
Then: 
\begin{equation}
H=\int_0^\infty dE\,E\,|E\rangle \langle E|  \label{2.1}
\end{equation}
being $|E\rangle $ ($\langle E|$) generalized right (left) eigenvectors of $%
H $ with eigenvalue $E$ \cite{I}\cite{I2}, and $dE$ the Lebesgue measure on
the real line.

The time evolution of a pure state is given by 
\begin{eqnarray}
|\Psi _t\rangle &=&e^{-iHt}\,|\Psi _0\rangle =\int_0^\infty dE\,|E\rangle
\,\Psi _t(E)  \nonumber \\
\Psi _t(E) &=&e^{-iEt}\,\langle E|\Psi _0\rangle .  \label{2.2}
\end{eqnarray}

The wave function $\Psi _t(E)$ has an oscillating time dependence with no
well defined limit for $t\rightarrow \infty $. However, it is possible to
obtain a well defined limit for the mean value of any observable represented
by an operator of the form 
\begin{eqnarray}
O &=&O^d+O^c  \nonumber \\
O^d &=&\int_0^\infty dE\,O_E\,|E\rangle \langle E|  \label{2.3} \\
O^c &=&\int_0^\infty dE\int_0^\infty dE^{\prime }\,O_{E\,E^{\prime
}}\,|E\rangle \langle E^{\prime }|.  \nonumber
\end{eqnarray}

We shall assume that $O^c$ (where the letter $c$ relates to correlations),
belongs to ${\cal O}^c,$ the Hilbert-Schmidt class (H-S) of operators,
including its own Hilbertian topology. This implies that $O^c$ is an
integral operator with a square integrable kernel function \cite{H}.
Alternatively, we could choose its weak closure in the space of all bounded
operators of the Hilbert space of states, ${\cal U}({\cal O}^c)$ which is a
von Neumann algebra (a factor of type I, having a countable relative
dimension \cite{Nai}) containing H-S \cite{Dix}. But here our aim is to deal
with the internal time superoperator, whose classical theory is formalized
in a Hilbertian language, and so we prefer to choose the H-S class.

In addition, we shall explicitly include a ''diagonal part'' $O^d$, of the
form indicated by (\ref{2.3}). $O^d$ will belong to ${\cal O}^d,$ the
maximal Abelian von Neumann algebra \cite{Dix} generated by the spectral
projections of a complete set of commuting observables of the system. This
send us outside of the H-S class \cite{prep}\cite{B}\cite{C}.

For operators representing observables, we shall assume the reasonable
non-Hilbertian condition of ''self adjointness'', i.e. $O_E=O_E^{*}$ and $%
O_{E\,E^{\prime }}=O_{E^{\prime }\,E}^{*}$ . The involutive Banach algebra
with identity \cite{Dix} of all possible operators $O=O^d+O^c$ will be
denoted by ${\cal O=O}^d\bigoplus {\cal O}^c$, and its involution by $%
\dagger $. (The sum is a direct one, because a non null diagonal operator
cannot be compact \cite{prep}, and all the H-S operators are compact \cite{H}%
. If we choose ${\cal U}({\cal O}^c),$ as it contains the identity, the sum
is not direct)

With the pure state (\ref{2.2}) we can construct the corresponding nuclear
(and then, of Hilbert-Smith class \cite{H}\cite{I}) density operator 
\begin{eqnarray}
\stackrel{\wedge }{\rho }_t &=&|\Psi _t\rangle \langle \Psi _t|=\int \int
dE\,dE^{\prime }\,|E\rangle \langle E^{\prime }|\,e^{-i\,(E-E^{\prime
})\,t}\,\langle E|\Psi _0\rangle \langle \Psi _0|E^{\prime }\rangle
\label{2.4} \\
Tr\,(\stackrel{\wedge }{\rho }_t) &=&Tr\,(\stackrel{\wedge }{\rho }%
_0)=\langle \Psi _0|\Psi _0\rangle =1.  \label{2.5}
\end{eqnarray}

In the usual quantum formalism, the mean value of an observable $O$ is 
\[
\langle O\rangle _t=Tr(\stackrel{\wedge }{\rho }_tO). 
\]

For an observable $O$ of the form given in (\ref{2.3}) we obtain: 
\begin{eqnarray}
\langle O\rangle _t &=&\int dE\,\langle E|\Psi _0\rangle \langle \Psi
_0|E\rangle \,O_E+  \nonumber \\
&&+\int \int dE\,dE^{\prime }\,e^{-i\,(E-E^{\prime })\,t}\,\langle E|\Psi
_0\rangle \langle \Psi _0|E^{\prime }\rangle \,O_{E\,E^{\prime }}
\label{2.6}
\end{eqnarray}

If the function $f(E,E^{\prime }):=\langle E|\Psi _0\rangle \langle \Psi
_0|E^{\prime }\rangle \,O_{E\,E^{\prime }}$ is integrable on $[0,\infty
)\times [0,\infty )$, the second term in (\ref{2.6}) goes to zero when $%
t\rightarrow \infty $, as a consequence of the Riemann-Lebesgue Lemma, and
we obtain 
\begin{equation}
\lim_{t\rightarrow \infty }\,\langle O\rangle _t=\int dE\,\langle E|\Psi
_0\rangle \langle \Psi _0|E\rangle \,O_E.  \label{2.7}
\end{equation}

We may try to find a density operator $\stackrel{\wedge }{\rho }_\infty
=\int \int dEdE^{\prime }(\stackrel{\wedge }{\rho }_\infty )_{E\,E^{\prime
}}|E\rangle \langle E^{\prime }|$ such that 
\[
\lim_{t\rightarrow \infty }\langle O\rangle _t=Tr(\stackrel{\wedge }{\rho }%
_\infty O)\int dE\,(\stackrel{\wedge }{\rho }_\infty )_{E\,E}O_E+\int \int
dE\,dE^{\prime }\,(\stackrel{\wedge }{\rho }_\infty )_{E\,E^{\prime
}}O_{E\,E^{\prime }} 
\]
which would imply 
\[
\int dE\,\langle E|\Psi _0\rangle \langle \Psi _0|E\rangle \,O_E=\int dE\,(%
\stackrel{\wedge }{\rho }_\infty )_{E\,E}O_E+\int \int dE\,dE^{\prime }\,(%
\stackrel{\wedge }{\rho }_\infty )_{E\,E^{\prime }}O_{E\,E^{\prime }} 
\]

But there is no function $(\stackrel{\wedge }{\rho }_\infty )_{E\,E^{\prime
}}$ satisfying the previous equation for arbitrary functions $O_E$ and $%
O_{E\,E^{\prime }}$, and therefore $\stackrel{\wedge }{\rho }_\infty $ do
not exist.

Moreover, for the pure state corresponding to a well defined value $E$ of
the energy and represented by the {\em generalized vector} $|E\rangle $,
expressions like $\langle E|E\rangle $ or $\langle E|H|E\rangle $ are not
defined.

These difficulties can be overcame by an extended definition of states as
functionals acting on the operators representing observables. This approach
was developed by I.Antoniu et al \cite{B}\cite{C}. If the observable is
represented by a ''self adjoint'' operator $O$ having diagonal part, as it
is the case for expression (\ref{2.3}), the state $\rho $ of the system is
represented by two functions $\rho _E$ and $\rho _{E\,E^{\prime }}$ such
that the mean value $\langle O\rangle _\rho $ of the observable $O$ in the
state $\rho $ is given by 
\begin{equation}
\langle O\rangle _\rho =\int dE\,\rho _E^{*}\,O_E+\int \int dE\,dE^{\prime
}\,\rho _{E\,E^{\prime }}^{*}\,O_{E\,E^{\prime }}.  \label{2.8}
\end{equation}

The mean value $\langle O\rangle _\rho $ is real if 
\begin{equation}
(\rho _E)^{*}=\rho _E,\qquad (\rho _{E\,E^{\prime }})^{*}=\rho _{E^{\prime
}\,E}.  \label{2.9}
\end{equation}

If $I$ denotes the identity operator, we also have: $\left\langle
I\right\rangle _\rho =\left\langle \int dE\,|E\rangle \langle
E|\right\rangle _\rho =1$ if 
\begin{equation}
\int_0^\infty dE\,\rho _E=1.  \label{2.10}
\end{equation}

Therefore, the states $\rho $ are represented by functionals acting on the
space of operators representing observables. The states can be expressed in
terms of the functionals $(E|$ and $(E\,E^{\prime }|$, defined by the
following relations 
\begin{equation}
(E|O):=O_E,\qquad (EE^{\prime }|O):=O_{EE^{\prime }}.  \label{2.10'}
\end{equation}

From (\ref{2.10'}) we can deduce the following formal relations 
\begin{eqnarray}
(E|E^{\prime }) &=&\delta (E-E^{\prime })  \nonumber \\
(E\,E^{\prime }|E^{\prime \prime }\,E^{\prime \prime \prime }) &=&\delta
(E-E^{\prime \prime })\delta (E^{\prime }-E^{\prime \prime \prime }) 
\nonumber \\
(E|E^{\prime }\,E^{\prime \prime }) &=&0  \nonumber \\
(E\,E^{\prime }|E^{\prime \prime }) &=&0,  \label{2.11}
\end{eqnarray}
where 
\begin{equation}
|E):=|E\rangle \langle E|\qquad |E\,E^{\prime }):=|E\rangle \langle
E^{\prime }|.  \label{2.12}
\end{equation}

Using (\ref{2.12}), we can give the following expression for an operator $%
|O) $ representing an observable with diagonal part 
\begin{equation}
|O)=\int dE\,O_E\,|E)+\int \int dE\,dE^{\prime }\,O_{E\,E^{\prime
}}\,|E\,E^{\prime }).  \label{2.13}
\end{equation}

Using (\ref{2.11}), the following expression can be given for a functional
representing a state $\rho $%
\[
(\rho |\in {\cal S}={\cal S}^d\bigoplus {\cal S}^c\subset
(as\,a\,convex\,subset)\,{\cal O}^{\prime }=\left( {\cal O}^d\right)
^{\prime }\bigoplus \left( {\cal O}^c\right) ^{\prime } 
\]
\begin{eqnarray}
(\rho | &=&(\rho ^d|+(\rho ^c|\quad \in {\cal S}  \nonumber \\
(\rho ^d| &=&\int dE\,\rho _E^{*}\,(E|\quad \in {\cal S}^d  \label{2.14} \\
(\rho ^c| &=&\int \int dE\,dE^{\prime }\,\rho _{E\,E^{\prime
}}^{*}\,(E\,E^{\prime }|\quad \in {\cal S}^c.  \nonumber
\end{eqnarray}

Using (\ref{2.11}), (\ref{2.13}) and (\ref{2.14}) we can easily prove that 
\begin{equation}
\langle O\rangle _\rho =(\rho |O).  \label{2.15}
\end{equation}

Conditions (\ref{2.9}) and (\ref{2.10}) can be written as 
\begin{eqnarray}
(\rho |O) &=&(\rho |O)^{*}\quad if\quad O^{\dagger }=O  \label{2.16} \\
(\rho |I) &=&(\rho |\int dE\,|E)=1.  \label{2.17}
\end{eqnarray}

Expression (\ref{2.17}) is a generalization of the concept of trace.

It is interesting to point out that {\em the formalism defined above already
contain the usual approach of quantum mechanics}{\bf . }In fact, consider a
pure state, represented by a normalized vector 
\begin{eqnarray}
|\Psi \rangle &=&\int dE\,\Psi (E)\,|E\rangle  \label{2.18} \\
\langle \Psi |\Psi \rangle &=&\int dE\,\Psi (E)^{*}\,\Psi (E)=1.
\label{2.19}
\end{eqnarray}

For an observable $O$ with diagonal singularity, as in (\ref{2.3}), using
the standard formalism we obtain 
\begin{eqnarray}
\langle O\rangle _\Psi &=&\langle \Psi |O|\Psi \rangle =\int dE\,\Psi
(E)^{*}\,\Psi (E)\,O_E+  \nonumber \\
&&\int \int dE\,dE^{\prime }\Psi (E)^{*}\,\Psi (E^{\prime
})\,O_{E\,E^{\prime }},  \label{2.20}
\end{eqnarray}

In the new formalism, a pure state is represented by the functional 
\begin{equation}
(\rho _{pure}|:=\int dE\,\Psi (E)^{*}\,\Psi (E)\,(E|+\int \int
dE\,dE^{\prime }\Psi (E)^{*}\,\Psi (E^{\prime })\,(E\,E^{\prime }|
\label{2.21}
\end{equation}

It is easy to verify from the definition (\ref{2.21}) that $(\rho
_{pure})_E^{*}=(\rho _{pure})_E$ and $(\rho _{pure})_{E\,E^{\prime
}}^{*}=(\rho _{pure})_{E^{\prime }\,E}$, and therefore $(\rho _{pure}|$
satisfies (\ref{2.16}). Condition (\ref{2.17}) is also verified by $(\rho
_{pure}|$ as a consequence of the normalization (\ref{2.19}) of the vector $%
|\Psi \rangle $. By acting with the functional $(\rho _{pure}|$ on $|O)$ we
obtain 
\[
(\rho _{pure}|O)=\langle \Psi |O|\Psi \rangle . 
\]

If the state is a mixture, represented in the standard formalism by a
density operator $\stackrel{\wedge }{\rho }$ satisfying 
\begin{eqnarray*}
\stackrel{\wedge }{\rho } &=&\int \int dE\,dE^{\prime }\,(\stackrel{\wedge }{%
\rho })_{E\,E^{\prime }}\,|E\rangle \langle E^{\prime }| \\
(\stackrel{\wedge }{\rho })_{E\,E^{\prime }}^{*} &=&(\stackrel{\wedge }{\rho 
})_{E^{\prime }\,E} \\
Tr(\stackrel{\wedge }{\rho }) &=&\int dE\,(\stackrel{\wedge }{\rho }%
)_{E\,E}=1,
\end{eqnarray*}
the mean value of an observable $O$ is given by 
\[
\langle O\rangle _{\stackrel{\wedge }{\rho }}=Tr(\stackrel{\wedge }{\rho }%
\,O)=\int dE\,(\stackrel{\wedge }{\rho })_{E\,E}O_E+\int \int dE\,dE^{\prime
}\,(\stackrel{\wedge }{\rho })_{E\,E^{\prime }}\,O_{E\,E^{\prime }}. 
\]

The mixture can also be described using the extended formalism, provided we
define 
\begin{equation}
(\rho _{mix}|:=\int dE\,(\stackrel{\wedge }{\rho })_{E\,E}\,(E|+\int \int
dE\,dE^{\prime }\,(\stackrel{\wedge }{\rho })_{E\,E^{\prime
}}\,(E\,E^{\prime }|.  \label{2.22}
\end{equation}

It is easy to verify 
\[
(\rho _{mix}|O)=Tr(\stackrel{\wedge }{\rho }\,O)\qquad (\rho
_{mix}|I)=1\qquad (\rho _{mix}|O)=(\rho _{mix}|O)^{*}. 
\]

As we see from (\ref{2.21}) and (\ref{2.22}), ordinary states (pure or
mixtures) satisfy $\rho _E=\rho _{E\,E}$, and therefore the diagonal and the
off-diagonal parts of $(\rho _{pure}|$ or $(\rho _{mix}|$ are not
independent. However, {\em the generalized formalism allows more general
states} represented by functionals $(\rho |$ satisfying (\ref{2.16}) and (%
\ref{2.17}) for which $\rho _E\neq \rho _{E\,E}$, i.e. states {\em which can
not be represented by normalized vectors or by density operators}{\bf .}

Consider for example a pure state corresponding to a well defined value $E$
of the energy. If we represent this state by the generalized eigenvector $%
|E\rangle $ of the Hamiltonian, $\langle E|E\rangle $ and $\langle
E|H|E\rangle $ are not defined. The standard trick is to make the spectrum
of the Hamiltonian discrete by putting the system in a box, and to make the
volume of the box very big after all the relevant calculations.

This trick is not necessary in the generalized formalism. Consider for
example the state $(E|$, for which the generalized trace is well defined 
\[
(E|I)=(E|\int dE^{\prime }|E^{\prime })=\int dE^{\prime }\,\delta
(E-E^{\prime })=1. 
\]

This state satisfy 
\[
\langle H^n\rangle =(E|H^n)=(E|\int dE^{\prime }|E^{\prime })\,(E^{\prime
})^n=\int dE^{\prime }\,\delta (E-E^{\prime })\,(E^{\prime })^n=E^n 
\]

from which we easily deduce that $(E|$ has a well defined value of the
energy 
\[
\langle H\rangle =(E|H)=E\qquad \langle (H-\langle H\rangle )^n\rangle
=(E|(H-\langle H\rangle )^n)=0. 
\]
Therefore $(E|$ represent a ''generalized pure state'' with energy $E$.

Concerning the time evolution of the states, it is determined by 
\begin{eqnarray}
{\Bbb U}_t\text{\ acting\ on\ }\rho \text{\ by}\;\;\;\;\;\;\;(\rho _t|O) &=&(%
{\Bbb U}_t\,\rho _0|O):=(\rho _0|{\Bbb U}_t^{+}\,O)  \nonumber \\
&:&=(\rho _0|e^{i\,{\Bbb L}^{+}\,t}\,O):=(\rho _0|e^{i\,H\,t}Oe^{-i\,H\,t}),
\label{2.23}
\end{eqnarray}
which also gives the relation between Schr\"{o}dinger and Heisemberg
pictures. The last equation is a special case of the general rule $({\Bbb M}%
\rho |O):=(\rho |{\Bbb M}^{+}O)$, which defines ${\Bbb M}$ acting on $\rho $
in terms of a certain ${\Bbb M}^{+}$ (that will be given in each case)
acting on $O$. Notice that this amount to give a generalized non-Hilbertian
''adjoint'' relation $+$, a kind of ''duality''.

The generalized Liouville-Von Neumann equation is 
\begin{eqnarray}
-i\frac d{dt}(\rho _t| &=&({\Bbb L}\,\rho _t|=(\rho _t|\,{\Bbb L}^{+} 
\nonumber \\
{\Bbb L}^{+}\,O &:&=H\,O-O\,H,  \label{2.24}
\end{eqnarray}
with the general solution 
\begin{eqnarray}
(\rho _t| &=&\int dE\,(\rho _t)_E^{*}\,(E|+\int \int dE\,dE^{\prime }\,(\rho
_t)_{E\,E^{\prime }}^{*}\,(E\,E^{\prime }|=  \nonumber \\
&=&\int dE\,(\rho _0)_E^{*}\,(E|+\int \int dE\,dE^{\prime }\,(\rho
_0)_{E\,E^{\prime }}^{*}\,e^{i\,(E-E^{\prime })\,t}\,(E\,E^{\prime }|.
\label{2.25}
\end{eqnarray}

Therefore 
\begin{equation}
\langle O\rangle _{\rho _t}=\int dE\,(\rho _0)_E^{*}\,O_E+\int \int
dE\,dE^{\prime }\,(\rho _0)_{E\,E^{\prime }}^{*}\,e^{i\,(E-E^{\prime
})\,t}\,O_{E\,E^{\prime }}.  \label{2.26}
\end{equation}

For integrable functions $f(E,E^{\prime })=(\rho _0)_{E\,E^{\prime
}}\,O_{E\,E^{\prime }}$, the second term of the previous expression vanishes
when $t\rightarrow \infty $, and therefore we have the weak limit 
\begin{equation}
\text{w-}\lim_{t\rightarrow \infty }(\rho _t|=\int dE\,(\rho _0)_E^{*}\,(E|
\label{2.27}
\end{equation}

The existence of weak limits when $t\rightarrow \infty $ makes this
formalism specially suitable to describe time evolution of decaying quantum
systems and the approach to equilibrium in statistical mechanics.

Up to now we expressed the spectral resolution of the operators representing
observables in terms of $|E):=|E\rangle \langle E|$ and $|EE^{\prime
}):=|E\rangle \langle E^{\prime }|$, being $|E\rangle $ ($\langle E|$)
generalized right (left) eigenvectors of the total Hamiltonian $H$ of the
system. Equations (2.11) define the corresponding functionals $(E|$ and $%
(E\,E^{\prime }|$ to expand the states. These generalized states and
observables are left and right generalized eigenvectors of the Liouville-Von
Neumann superoperator ${\Bbb L}^{+}$%
\begin{eqnarray*}
(E|{\Bbb L}^{+} &=&0,\qquad (E\,E^{\prime }|{\Bbb L}^{+}=(E-E^{\prime
})(E\,E^{\prime }| \\
{\Bbb L}^{+}|E) &=&0,\qquad {\Bbb L}^{+}|EE^{\prime })=(E-E^{\prime
})|EE^{\prime })
\end{eqnarray*}

The superoperator ${\Bbb L}^{+}$can be written as 
\begin{equation}
{\Bbb L}^{+}=\int \int dE\,dE^{\prime }\,(E-E^{\prime })\,|EE^{\prime
})(E\,E^{\prime }|  \label{2.28}
\end{equation}

We can make the change of variables 
\begin{equation}
\nu =E-E^{\prime },\qquad \lambda =\frac{E+E^{\prime }}2,  \label{2.29}
\end{equation}
and define 
\begin{eqnarray}
|\lambda \nu ) &:&=|EE^{\prime })=|\lambda +\frac \nu 2\rangle \langle
\lambda -\frac \nu 2|  \nonumber \\
|\lambda ) &:&=|E)=|\lambda \rangle \langle \lambda |  \nonumber \\
(\lambda \nu | &:&=(EE^{\prime }|  \nonumber \\
(\lambda | &:&=(E|.  \label{2.30}
\end{eqnarray}

Using (\ref{2.11}) and (\ref{2.30}) we obtain 
\begin{eqnarray}
(\lambda |\lambda ^{\prime }) &=&\delta (\lambda -\lambda ^{\prime }) 
\nonumber \\
(\lambda |\lambda ^{\prime }\nu ^{\prime }) &=&0  \nonumber \\
(\lambda \nu |\lambda ^{\prime }\nu ^{\prime }) &=&\delta (\lambda -\lambda
^{\prime })\delta (\nu -\nu ^{\prime })  \nonumber \\
(\lambda \nu |\lambda ^{\prime }) &=&0.  \label{2.31}
\end{eqnarray}

For the superoperators ${\Bbb L}^{+}$ and ${\Bbb I}^{+}$ (defined by ${\Bbb I%
}^{+}O=O$ for all $O\in {\cal O}$), we obtain 
\begin{equation}
{\Bbb L}^{+}=\int_0^\infty d\lambda \int_{-2\lambda }^{2\lambda }d\nu \,\nu
\,|\lambda \nu )(\lambda \nu |  \label{2.32}
\end{equation}
\begin{eqnarray}
{\Bbb I}^{+} &=&{\Bbb I}_d^{+}+{\Bbb I}_c^{+}  \nonumber \\
{\Bbb I}_d^{+} &=&\int_0^\infty dE\,|E)\,(E|=\int_0^\infty d\lambda
\,|\lambda )(\lambda |  \nonumber \\
{\Bbb I}_c^{+} &=&\int \int dE\,dE^{\prime }\,|EE^{\prime })\,(E\,E^{\prime
}|=\int_0^\infty d\lambda \int_{-2\lambda }^{2\lambda }d\nu \,|\lambda \nu
)(\lambda \nu |  \label{2.33}
\end{eqnarray}

\section{The time superoperator}

In classical context, the time superoperator $T$ is defined through $%
[T,L]=i\,I$, where $I$ stand for the identity superoperator acting on
fluctuations of the equilibrium state $\rho _\infty $ \cite{D}\cite{E}\cite
{F}. This amount of projecting on the off diagonal part of the state,
because in that case the ''diagonal part'' is trivial: ${\Bbb C}\rho _\infty 
$, being ${\Bbb C}$ the complex number field. Therefore, in the quantum
case, where we have a more general diagonal part, we propose the following
definition: 
\begin{equation}
\lbrack {\Bbb T},{\Bbb L}]=i\,{\Bbb I}_c,  \label{3.0}
\end{equation}
where ${\Bbb T}$, ${\Bbb L}$ and ${\Bbb I}_c$ are superoperators acting on
the states. ${\Bbb L}$ is the generator of time evolution for the states,
and ${\Bbb I}_c$ is the projection onto the off diagonal part of the states.

Our hypotesis on the spectrum of $H$, determines the corresponding spectral
properties of ${\Bbb L}$ acting on $\left( {\cal O}^c\right) ^{\prime
}\equiv {\cal O}^c\supset {\cal S}^c.$ In fact, ${\Bbb L}$ has uniform
Lebesgue spectrum on all the real line. It is a well known fact \cite{J}\cite
{K}\cite{F} that this implies the existence of a spectral measure ${\Bbb E}$
from the Lebesgue $\sigma $-algebra of ${\Bbb R}$ to the projection
superoperators on the Hilbert-Schmidt class ${\cal O}^c$, such that $\left( 
{\Bbb U}_t\mid _{\left( {\cal O}^c\right) ^{\prime }},{\Bbb E}\right) $ is a
system of imprimitivity based on ${\Bbb R}$, that is to say, for every
Lebesgue-measurable set $\Delta $, we have: 
\begin{equation}
{\Bbb U}_{-t}{\Bbb E(}\Delta {\Bbb )U}_t\mid _{\left( {\cal O}^c\right)
^{\prime }}={\Bbb E(}\Delta {\Bbb +\,}t)  \label{3.0'}
\end{equation}

Consequently there exists ${\Bbb T}\mid _{\left( {\cal O}^c\right) ^{\prime
}}{\Bbb =}\int_{-\infty }^{+\infty }s\,d{\Bbb E}$. Then we can define ${\Bbb %
T}$ over ${\cal O}^{\prime }$ by requiring:${\Bbb T}\mid _{\left( {\cal O}%
^d\right) ^{\prime }}=0$ . If we also define ${\Bbb E(}\Delta {\Bbb )}\mid
_{\left( {\cal O}^d\right) ^{\prime }}=0$, then $\int_{-\infty }^{+\infty
}s\,d{\Bbb E}\mid _{\left( {\cal O}^d\right) ^{\prime }}=0$, and $%
\int_{-\infty }^{+\infty }d{\Bbb E}\mid _{\left( {\cal O}^d\right) ^{\prime
}}=0$, and we can write 
\begin{eqnarray}
{\Bbb T} &=&\int_{-\infty }^{+\infty }s\,d{\Bbb E}  \label{3.00+} \\
{\Bbb I}_c &=&\int_{-\infty }^{+\infty }d{\Bbb E}  \nonumber \\
{\Bbb U}_{-t}{\Bbb E(}\Delta {\Bbb )U}_t &=&{\Bbb E(}\Delta {\Bbb +\,}t)
\label{3.00*}
\end{eqnarray}

If we define ${\Bbb E}_s:={\Bbb E}(-\infty ,s],$ the previous expresions
give:

\begin{eqnarray}
{\Bbb T} &=&\int_{-\infty }^{+\infty }s\,d{\Bbb E}_s  \label{3.0+} \\
{\Bbb I}_c &=&\int_{-\infty }^{+\infty }d{\Bbb E}_s  \nonumber \\
{\Bbb U}_{-t}{\Bbb E}_s{\Bbb U}_t &=&{\Bbb E}_{s-t}  \label{3.0*}
\end{eqnarray}

In what follows we are going to obtain explicit expressions for ${\Bbb T}$
and ${\Bbb E}$ starting from the generalized spectral decomposition of the
Hamiltonian given in equation (\ref{2.1}).

Using the relations 
\begin{eqnarray*}
({\Bbb L}\rho |O) &=&(\rho |{\Bbb L}^{+}O):=(\rho |[H,O]) \\
({\Bbb I}_c\rho |O) &=&(\rho |{\Bbb I}_c^{+}O):=(\rho |O^c) \\
({\Bbb T}\rho |O) &=&(\rho |{\Bbb T}^{+}O)
\end{eqnarray*}
we obtain 
\begin{equation}
\lbrack {\Bbb T}^{+},{\Bbb L}^{+}]=i\,{\Bbb I}_c^{+},  \label{3.1}
\end{equation}
where ${\Bbb L}^{+}$ is given by (\ref{2.32}) or (\ref{2.28}), and 
\[
{\Bbb I}_c^{+}=\int_0^\infty d\lambda \int_{-2\lambda }^{2\lambda }d\nu
\,|\lambda \nu )(\lambda \nu | 
\]

From (\ref{3.1}) we obtain 
\begin{eqnarray}
(\nu -\nu ^{\prime })(\lambda ^{\prime }\nu ^{\prime }|{\Bbb T}^{+}|\lambda
\nu ) &=&i\,\delta (\lambda ^{\prime }-\lambda )\delta (\nu ^{\prime }-\nu )
\nonumber \\
\nu (\lambda ^{\prime }|{\Bbb T}^{+}|\lambda \nu ) &=&0  \nonumber \\
\nu ^{\prime }(\lambda ^{\prime }\nu ^{\prime }|{\Bbb T}^{+}|\lambda ) &=&0
\label{3.2}
\end{eqnarray}

A possible solution for equations (\ref{3.2}) is 
\begin{eqnarray}
(\lambda ^{\prime }\nu ^{\prime }|{\Bbb T}^{+}|\lambda \nu ) &=&i\,\delta
(\lambda ^{\prime }-\lambda )\delta ^{\prime }(\nu ^{\prime }-\nu ) 
\nonumber \\
(\lambda ^{\prime }|{\Bbb T}^{+}|\lambda \nu ) &=&0  \nonumber \\
(\lambda ^{\prime }\nu ^{\prime }|{\Bbb T}^{+}|\lambda ) &=&0  \nonumber \\
(\lambda ^{\prime }|{\Bbb T}^{+}|\lambda ) &=&0  \label{3.3}
\end{eqnarray}
and therefore 
\[
(\rho |{\Bbb T}^{+}|O)=i\int_0^\infty d\lambda \int_{-2\lambda }^{2\lambda
}d\nu \,(\rho |\lambda \nu )\frac \partial {\partial \nu }(\lambda \nu |O). 
\]

To obtain the right eigenvectors $|\varphi )$ of the time superoperator we
make a Fourier expansion on the $\nu $ variable 
\begin{eqnarray*}
|\varphi ) &=&\int_0^\infty d\lambda \,|\lambda )(\lambda |\varphi
)+\int_0^\infty d\lambda \int_{-2\lambda }^{2\lambda }d\nu \,|\lambda \nu
)(\lambda \nu |\varphi ) \\
&=&\int_0^\infty d\lambda \,|\lambda )\,\varphi (\lambda )+\int_0^\infty
d\lambda \int_{-2\lambda }^{2\lambda }d\nu \,|\lambda \nu )\sum_{n=-\infty
}^{+\infty }\varphi _n(\lambda )\exp (-\frac{in\pi \nu }{2\lambda }).
\end{eqnarray*}

Replacing this last expression in ${\Bbb T}^{+}|\varphi )=s\,|\varphi )$ we
obtain the following generalized eigenvalues and eigenvectors 
\begin{equation}
\begin{array}{ll}
s=0 & \quad |\varphi _\lambda ):=|\lambda ) \\ 
s=0 & \quad |\varphi _{0\lambda }):=\frac 1{2\sqrt{\lambda }}\int_{-2\lambda
}^{+2\lambda }d\nu \,|\lambda \nu ) \\ 
s\neq 0 & \quad |\varphi _{sn}):=\frac 1{2\sqrt{|s|}}\int_{-\frac{n\pi }s}^{%
\frac{n\pi }s}d\nu \,|\frac{n\pi }{2s},\nu )\exp (-is\nu ).
\end{array}
\label{3.4}
\end{equation}

The generalized left eigenvectors, satisfying $(\varphi |{\Bbb T}%
^{+}=s\,(\varphi |$, are 
\begin{equation}
\begin{array}{ll}
s=0 & \quad (\varphi _\lambda |:=(\lambda | \\ 
s=0 & \quad (\varphi _{0\lambda }|:=\frac 1{2\sqrt{\lambda }}\int_{-2\lambda
}^{+2\lambda }d\nu \,(\lambda \nu | \\ 
s\neq 0 & \quad (\varphi _{sn}|:=\frac 1{2\sqrt{|s|}}\int_{-\frac{n\pi }s}^{%
\frac{n\pi }s}d\nu \,(\frac{n\pi }{2s},\nu |\exp (is\nu ).
\end{array}
\label{3.5}
\end{equation}

In the expressions (\ref{3.4}) and (\ref{3.5}), $n$ is an integer positive
(negative) number if $s>0$ ($s<0$). These generalized eigenvectors form a
biorthonormal system 
\begin{eqnarray}
(\varphi _\lambda |\varphi _{\lambda ^{\prime }}) &=&\delta (\lambda
-\lambda ^{\prime })  \nonumber \\
(\varphi _{0\lambda }|\varphi _{0\lambda ^{\prime }}) &=&\delta (\lambda
-\lambda ^{\prime })  \nonumber \\
(\varphi _{sn}|\varphi _{s^{\prime }n^{\prime }}) &=&\delta (s-s^{\prime
})\delta _{nn^{\prime }}  \nonumber \\
(\varphi _{sn}|\varphi _{0\lambda ^{\prime }}) &=&(\varphi _{sn}|\varphi
_{\lambda ^{\prime }})=(\varphi _{0\lambda }|\varphi _{\lambda ^{\prime }})=0
\label{3.6}
\end{eqnarray}

This system is also complete 
\begin{eqnarray}
{\Bbb I}^{+} &=&\int\limits_0^{+\infty }d\lambda \,|\varphi _\lambda
)(\varphi _\lambda |+\int\limits_{-\infty }^{0^{-}}ds\sum\limits_{n=-\infty
}^{-1}|\varphi _{sn})(\varphi _{sn}|+  \nonumber \\
&&+\int\limits_0^{+\infty }d\lambda \,|\varphi _{0\lambda })(\varphi
_{0\lambda }|+\int\limits_{0^{+}}^{+\infty }ds\sum\limits_{n=1}^{+\infty
}|\varphi _{sn})(\varphi _{sn}|.  \label{3.7}
\end{eqnarray}

The spectral measure ${\Bbb E}_s$, in weak sense, is: 
\begin{equation}
\begin{tabular}{l}
${\Bbb E}_s:=\int\limits_{-\infty }^sds^{\prime }\sum\limits_{n=-\infty
}^{-1}|\varphi _{s^{\prime }n})(\varphi _{s^{\prime }n}|\quad \qquad s<0$ \\ 
${\Bbb E}_s:=\int\limits_{-\infty }^{0^{-}}ds^{\prime
}\sum\limits_{n=-\infty }^{-1}|\varphi _{s^{\prime }n})(\varphi _{s^{\prime
}n}|+\int\limits_0^{+\infty }d\lambda \,|\varphi _{0\lambda })(\varphi
_{0\lambda }|+$ \\ 
$+\int\limits_{0^{+}}^sds^{\prime }\sum\limits_{n=1}^{+\infty }|\varphi
_{s^{\prime }n})(\varphi _{s^{\prime }n}|\quad s>0$%
\end{tabular}
\label{3.8}
\end{equation}
and therefore 
\begin{equation}
{\Bbb I}^{+}=\int\limits_0^{+\infty }d\lambda \,|\varphi _\lambda )(\varphi
_\lambda |+\int\limits_{-\infty }^{+\infty }d{\Bbb E}_s,\quad {\Bbb I}%
_c^{+}=\int\limits_{-\infty }^{+\infty }d{\Bbb E}_s,\quad {\Bbb T}%
^{+}=\int\limits_{-\infty }^{+\infty }s\,d{\Bbb E}_s.  \label{3.9}
\end{equation}

\section{Lyapounov variables}

Integrating the identity function $s$ in equation (\ref{3.0*}) with respect
to $s,$ we have 
\[
{\Bbb U}_{-t}\left( \int\limits_{-\infty }^{+\infty }s\,d{\Bbb E}_s\right) 
{\Bbb U}_t=\int\limits_{-\infty }^{+\infty }s\,d{\Bbb E}_{s-t} 
\]

With the change of variable $u=s-t,$ we obtain

\begin{equation}
{\Bbb U}_{-t}\,{\Bbb T\,U}_t={\Bbb T+}t\,{\Bbb I}_c,  \label{4.1}
\end{equation}
or the dual equation 
\begin{equation}
{\Bbb U}_t^{+}\,{\Bbb T}^{+}{\Bbb \,U}_{-t}^{+}={\Bbb T}^{+}{\Bbb +}t\,{\Bbb %
I}_c^{+}.  \label{4.2}
\end{equation}

The superoperator ${\Bbb A}({\Bbb T}^{+})$ (which will define the non
unitary similarity transformation, the ''$\Lambda $'' of the Brussels
group), is defined by 
\begin{equation}
{\Bbb A}({\Bbb T}^{+}):={\Bbb I}_d^{+}+A({\Bbb T}^{+})=\int\limits_0^{+%
\infty }d\lambda \,|\varphi _\lambda )(\varphi _\lambda
|+\int\limits_{-\infty }^{+\infty }A(s)\,d{\Bbb E}_s,  \label{4.3}
\end{equation}
where $A(s)$ is a positive real function satisfying $A(s_1)>A(s_2)$ for $%
s_1<s_2$. From (\ref{4.2}) and (\ref{4.3}) 
\begin{equation}
{\Bbb U}_t^{+}\,{\Bbb A}({\Bbb T}^{+}){\Bbb \,U}_{-t}^{+}=\int\limits_0^{+%
\infty }d\lambda \,|\varphi _\lambda )(\varphi _\lambda
|+\int\limits_{-\infty }^{+\infty }A(s+t)\,d{\Bbb E}_s.  \label{4.4}
\end{equation}

We also define the transformed states 
\begin{equation}
(\widetilde{\rho }|:=(\rho |{\Bbb A}({\Bbb T}^{+})  \label{4.5}
\end{equation}
and the Lyapounov variable\footnote{%
In section 2 we defined the correlation part of the states as functionals of
the H-S class, which is a Hilbert space, of the form 
\[
(\rho ^c|=\int \int dE\,dE^{\prime }\,\rho _{EE^{\prime }}^{*}(EE^{\prime
}|. 
\]
\par
By the Riesz representation theorem, there is a corresponding 'dual
operator' defined by 
\[
|\rho ^c)=\int \int dE\,dE^{\prime }\,\rho _{EE^{\prime }}|EE^{\prime }). 
\]
} 
\begin{equation}
L(t):=\left( \widetilde{\rho }_t^c|\widetilde{\rho }_t^c\right) _{H-S}=Tr(%
\widetilde{\rho }_t^c|\widetilde{\rho }_t^c)  \label{4.6}
\end{equation}

Using (\ref{4.4}) we deduce 
\begin{eqnarray}
L(t) &=&(\rho _0^c|{\Bbb U}_t^{+}\,{\Bbb A}^2({\Bbb T}^{+}){\Bbb \,U}%
_{-t}^{+}|\rho _0^c)_{H-S}=  \nonumber \\
&=&(\rho _0^c|\int\limits_{-\infty }^{+\infty }A^2(s+t)\,d{\Bbb E}_s|\rho
_0^c)_{H-S}  \label{4.7}
\end{eqnarray}

Being $A(s)$ a decreasing function, $L(t)$ is a positive decreasing function
of the time (measuring the decay of correlations), reaching the minimum
value for $t\rightarrow \infty $%
\begin{equation}
L_{\min }=\lim_{t\rightarrow \infty }L(t)=0.  \label{4.8}
\end{equation}

The minimum value of $L$ is independent of the initial state energy
distribution. However {\em the final state (}$t\rightarrow \infty ${\em ) is
not unique.} This is a consequence of the fact that we are considering the
time evolution of an isolated system in a state which is a mixture of
different energies. In this case there is no interaction capable to
rearrange the energy distribution, and the system keeps memory of the
initial condition.

However, the existence of the time superoperator and the Liapounov variable
reflects a special kind of instability and intrinsic irreversibility of
quantum systems with continuum spectrum.

\section{Generalized K-flows}

The classical concept of K-flow is generalized in \cite{G} to cover
situations encountered in nonequilibrium quantum statistical mechanics.

It is known that, given a faithful (injective) normal (\cite{Take}\cite{Dix}%
) state $\rho _\infty $ on a von Neumann algebra ${\cal U}$ (for example a
thermal equilibrium state), there exists a unique continuous one parameter
group $\sigma ({\Bbb R})$ of automorphism of ${\cal U}$ with respect to
which $\rho _\infty $ satisfies the KMS (Kubo-Martin-Schwinger) boundary
condition. This group is called the modular group canonically associated to $%
\rho _\infty .$ Now, a generalized K-flow is defined as a tetrad $({\cal U}%
,\rho _\infty ,\alpha ({\Bbb R}),{\cal A}),$ where ${\cal U}$ is a von
Neumann algebra acting on a separable Hilbert space ${\cal H};$ $\rho
_\infty $ is a faithful normal state on ${\cal U}$; $\alpha ({\Bbb R})$ is a
continuous one parameter group of automorphism of ${\cal U}$ such that $\rho
_\infty \circ \alpha (t)=\rho _\infty $ for every $t\in {\Bbb R}$; and $%
{\cal A}$, is a von Neumann subalgebra of ${\cal U}$ that is stable under
the modular group, and has ''the K-property'' \cite{Arn}\cite{corn}, that is
to say:

i) ${\cal A}\subset \alpha (t)[{\cal A}]$ for every $t\in {\Bbb R}^{+}$

ii) the von Neumann algebra generated by $\{\alpha (t)[{\cal A}]$ $:$ $t\in 
{\Bbb R}\}$ coincides with ${\cal U}$

iii) the largest von Neumann algebra contained in all $\alpha (t)[{\cal A}]$
for every $t\in {\Bbb R},$ is ${\Bbb C}I$

As it is well known, every von Neumann algebra is a C*-algebra, and
therefore they admit an essentialy unique cyclic representation $\pi :{\cal %
U\rightarrow B}({\cal H})$ into the bounded operators of a Hilbert space.
This is the so called ''GNS (Gel'fand-Naimark-Segal) Construction'' \cite
{bogo}. Let ${\Bbb L}$ be the generator of the strongly continuous one
parameter, unitary group ${\Bbb U}({\Bbb R})$ implementing (by the
representation $\pi )$ $\alpha ({\Bbb R})$ on ${\cal H}$ ,and let $\Phi
_\infty \in {\cal H}$ denotes the cyclic and separating element for ${\cal U}
$ such that 
\begin{eqnarray}
\rho _\infty [O] &=&\left( \Phi _\infty \mid \pi (O)\Phi _\infty \right) _{%
{\cal H}}\;for\;every\;O\in {\cal U}  \nonumber \\
{\Bbb U}(t)\Phi _\infty &=&\Phi _\infty \;for\;every\;t\in {\Bbb R}
\label{4.9}
\end{eqnarray}

Therefore $0$ is a simple eigenvalue of ${\Bbb L}$, and it has uniform
Lebesgue spectrum on $({\Bbb C}\Phi )^{\perp }$ \cite{G}.

There is a generalized Kolmogoroff entropy, and a proof that every non
singular generalized K-flow has strictly positive entropy, and is stongly
mixing, in the sense that, for every $O$ and $Q$ in ${\cal U}$ : 
\[
lim_{\left| t\right| \rightarrow \infty }\;\rho _\infty \left[ Q\cdot \alpha
\left( t\right) \left[ O\right] \right] =\rho _\infty \left[ Q\right] \rho
_\infty \left[ O\right] 
\]

Thus, the dynamical evolution in a von Neumann algebra of operators
belonging to a separable Hilbert space ${\cal H}$, can be splited into a
1-dimensional ''diagonal part'', equal (or isomorphic) to ${\Bbb C},$ plus
an ''off diagonal'' part. In this last one, the flow can be represented as a
strongly continuous unitary group of ${\cal H}$, {\em whose generator has
uniform Lebesgue spectrum}{\bf . }

The Brussels group has shown the existence of an Internal Time
Superoperator, as well as its associated similarity transformations and
Liapounov variables, for every classical K-flow. Those elements were used to
prove the intrinsic irreversibilty of these dynamical systems.

Now, {\em what we have done can be seen as the corresponding generalization
for the quantum case.} In fact, we can assume, in a ''Heisemberg picture'',
that our generalized evolution is taking place in the von Neumann algebra $%
{\cal U}={\Bbb C(}O_0^d\mid +\;{\cal U}({\cal O}^c)$ , because the canonical
inclusion 
\[
i:{\Bbb C(}O_0^d\mid \bigoplus {\cal O}^c\rightarrow {\Bbb C(}O_0^d\mid +\;%
{\cal U}({\cal O}^c) 
\]
is continuous. This, in turn, can be demostrated by the following argument.
Let $\{(a_n,T_n)\}$ , with 
\[
T_n(\psi )=\sum\limits_{k=1}^\infty \tau _{nk}<\Omega _k\mid \psi >\Omega
_k\;\;where\;\sum\limits_{k=1}^\infty \mid \tau _{nk}\mid ^2<\infty 
\]
be a sequence of the domain, coverging to $(o,0).$ This implies that $%
a_n\rightarrow o$ in ${\Bbb C},$ and $T_n\rightarrow 0$ in the H-S topology,
i.e.: $Tr(T_n^{\dagger }T_n)=\sum\limits_{k=1}^\infty \mid \tau _{nk}\mid
^2\rightarrow o.$ Then $T_n\rightarrow 0$ also in the weak operator
topology, which is the topology of ${\cal U}({\cal O}^c).$

\section{Conclusions}

We have shown {\em the intrinsic irreversibility of a class of quantum
systems with certain spectral properties, that includes some Large Poincare
Systems, the generalized quantum K-flows, and certainly, the free quantum
particle}, whose Hamiltonian satisfies the spectral hypothesis of our
result, and therefore goes to a diagonal final state, has an internal time
operator, etc., as we have demostrated. This is not difficult to understand,
because this system is very different to its classical analog. In fact, free
wave packets dispersion is a well known result. Additionally, in David
Bhom's ''ontological interpretation'' of quantum mechanics, that frequently
provides an heuristic picture of what is implied in the Schr\"{o}dinger
equation (independently of its own validity or not), even ''free'' wave
packets interact with a ''quantum potential'', so their equation of motion
is nonlinear and ''chaotic''\cite{L}\cite{M}.

\end{document}